\documentclass[prd,10pt, tightenlines, twocolumn, nofootinbib]{revtex4-1}
\input epsf.tex
\epsfclipon

\usepackage{slashed}
\usepackage{color}

\usepackage{epsfig}
\usepackage{epstopdf}
\usepackage{epsf}
\input{epsf.sty}
\usepackage{graphicx,amsmath,amssymb}
\usepackage{multirow}
\usepackage[normalem]{ulem}
\usepackage{bbm,bm,amsmath,amssymb}
\usepackage{float}
\usepackage{mathtools}
\usepackage[pdftex,bookmarks,linktocpage,pdfpagelabels,plainpages=false,hyperfigures,linkcolor=blue,citecolor=blue,urlcolor=blue]{hyperref}
\hypersetup{colorlinks=true}

\allowdisplaybreaks
 
\newcommand{\BF}{\mathbf{F}}
\newcommand{\BB}{\mathbf{B}}
\newcommand{\nn}{\nonumber}

\newcommand{\AppRef}[1]{Appendix~\ref{#1}}

\usepackage{comment}
\def\be{\begin{equation}}
\def\ee{\end{equation}}
\def\figs/B{B}
\def\bea{\begin{eqnarray}}
\def\eea{\end{eqnarray}}
\def\bg{\begin{eqnarray}}
\def\nd{\end{eqnarray}}
\def\sin{{\rm sin}}

\newcommand{\figref}[1]{Fig.~\ref{#1}}

\begin{document}

\title{Black hole and cosmological analysis of BF sequestered gravity} 
\author{Stephon Alexander, Steven J. Clark, Gabriel Herczeg, and Michael W. Toomey}
\affiliation{ Brown Theoretical Physics Center and Department of Physics, Brown University, 182 Hope Street, Providence, Rhode Island, 02903}

\date{\today}

\begin{abstract}
We study a minimal extension of a recently proposed modification of general relativity that draws on concepts from topological field theory to resolve aspects of the cosmological constant problem. In the original model, the field content of general relativity was augmented to include a gauge field and an adjoint-valued two-form without modifying the classical gravitational dynamics. Here we include a kinetic term for the gauge field which sources fluctuations of the cosmological constant. We then study spherically symmetric black holes and a simple  homogeneous, isotropic cosmological model predicted by the extended theory. For the black hole case, we observe deviations near the event horizon as well as a ``charge"-like behavior induced by the gauge field. In the cosmological case, $\dot{H}$ is always positive and some solutions asymptote to a constant $H$. 
\end{abstract}

\maketitle

\section{Introduction}

The discrepancy between the value of the cosmological constant inferred from measurements of the acceleration of the expansion rate of the universe and the value predicted for the zero-point energy by quantum field theory is famously enormous, with some estimates ranging as high as 120 orders of magnitude \cite{adler1995vacuum}. The chasm between these two values, known as the cosmological constant problem \cite{ weinberg1989cosmological, Zeldovich:1967gd, padilla2015lectures, Bousso2007TheCC},
has vexed theorists for decades. 

Over the past few years, a new discrepancy has emerged between measurements of the current expansion rate of the universe, with cosmic microwave background data from the early universe measured by the Planck satellite \cite{Planck:2018vyg} suggesting a significantly lower value for the Hubble parameter than the one inferred from measurements of nearby stars and galaxies made by the SH0ES collaboration \cite{Riess:2019cxk}, among others \cite{Huang:2019yhh, Wong:2019kwg, Pesce:2020xfe, LIGOScientific:2017adf, DES:2019ccw, DES:2017txv, Freedman:2020dne, Ivanov:2019hqk, DAmico:2019fhj, Troster:2019ean, Sandage:2006cv, 2013ApJS..208...19H, ACTPol:2016kmo}.
This ``Hubble tension" could be the result of some as-yet unidentified systematic error in the early or late measurements of the Hubble parameter, but it is also possible that both measurements are accurate, and the source of the discrepancy is due to unknown physics affecting the dynamical evolution of the universe. 

Given these discrepancies, now is an auspicious time to consider modifications of general relativity and the standard model that have the potential to resolve both issues. In this paper, we will focus on the first possibility by further developing a recently proposed modification of general relativity that draws on topological field theory to resolve aspects of the cosmological constant problem \cite{Alexander:2018tyf, alexander2020chiral}. Here, we explore a natural extension of this theory that has dramatic implications for cosmological dynamics. We review the relevant aspects of the original theory, define its extension, and study spherically symmetric black hole solutions and a simple homogeneous, isotropic cosmological model.

The black hole solutions predicted by the model have some interesting properties.
The dominant behavior at large $r$ mimics the (A)dS-Schwarzschild black hole, however subleading corrections of order $r^{2/3}$ are also present. Finally, the solution contains a term that is identical to the term proportional to the electric charge in a Reissner-Nordstr\"{o}m black hole. However, since we do not include an external current in our model, there is no notion of electric charge available in the theory. Consequently, all black holes in this theory have the same effective electric charge determined completely by the gravitational and gauge couplings (and possibly an additional integration constant related to the asymptotic value of the effective cosmological constant). In this sense, the black holes may be regarded as having unit charge, giving them the character of fundamental particles.

The homogeneous, isotropic cosmological model we study cannot be solved exactly, so we rely on numerical methods to analyze its behavior.
We find that the model admits solutions where the effective cosmological constant tends to a true constant at late times, which is consistent with our current dark-energy dominated cosmological epoch. While the early-time dynamics do not appear to be supported by observation, it is possible that a realistic picture of cosmological evolution could emerge after incorporating matter and radiation into the model.

\section{Review of BF-Coupled Gravity}

\label{BFgravity}

Here we review a model presented in \cite{Alexander:2018tyf} and developed further in \cite{alexander2020chiral} that will provide the background for our analysis. This model couples Einstein gravity to a topological field theory (BF theory) such that the volume form determined by the spacetime metric is constrained to be equal to a volume form constructed from one of the fields of the BF theory. First, we will introduce the basic aspects of  BF theory, and then review the BF-coupled gravity model and briefly discuss its implications for the cosmological constant problem.

\subsection{BF theory}
BF theory is a rich subject with interesting connections to gravity \cite{Plebanski:1977zz, PhysRevLett.38.739, Freidel_2012, Birmingham:1991ty, cattaneo1995topological, Witten:1988hc} and string theory \cite{Baez:2006sa, baez2007exotic, Brooks:1994nn}. We will not attempt to give a thorough introduction to the topic here, but merely highlight the aspects of the theory that are most relevant for our current considerations. For a detailed review, see for example \cite{cattaneo1995topological, Birmingham:1991ty,  Brooks:1994nn}.

Given an $n$-dimensional manifold $M$ and a semi-simple Lie group $G$, the action of BF theory is simply
\be
S_{BF} = \int_M \mbox{tr}(\mathbf{B}\wedge \mathbf{F}), \label{BF}
\ee
where $\mathbf{B}$ is an adjoint valued $(n-2)$-form, and 
\be
\mathbf{F} = {\rm d}\mathbf{A} + \mathbf{A} \wedge \mathbf{A}
\ee
 is the curvature two-form of a connection $\mathbf{A}$ taking values in the Lie algebra of $G$.

If $M$ has dimension three or four, the action can be supplemented with an additional term  whose strength is controlled by a coupling $\mu$ that is often call the ``cosmological constant." Since we are interested in BF theory defined over a physical spacetime, we restrict attention to the four-dimensional case, where one can include a term proportional to $\mathbf{B}\wedge \mathbf{B}$:
\be
S_4 = \int_M \mbox{tr}(\mathbf{B}\wedge \mathbf{F} + \frac{\mu}{2}\mathbf{B}\wedge \mathbf{B}). \label{BF4}
\ee
The equations of motion for $\mathbf{A}$ and $\mathbf{B}$ are then
\be
{\rm d}_{\mathbf{A}}\mathbf{B} = 0, \qquad \mathbf{F} + \mu\mathbf{B} = 0,
\ee
where ${\rm d}_{\mathbf{A}}\mathbf{Q} = {\rm d}\mathbf{Q} + [\mathbf{A}, \mathbf{Q}]$. 

The action is invariant under the usual gauge transformation
\be
\mathbf{A} \to g^{-1}\mathbf{A}g + g^{-1}{\rm d}g,
 \label{firstGauge}
\ee
provided that $\mathbf{B}$ transforms in the adjoint representation
\be
 \qquad \mathbf{B} \to g^{-1}\mathbf{B}g.
\ee
In addition to the usual gauge transformations, the action $S_4$ is also invariant under another set of transformations parametrized by a Lie algebra valued one-form $\boldsymbol{\eta}$:
\be
\mathbf{A} \to \mathbf{A} + \mu\boldsymbol{\eta}, \qquad  \mathbf{B} \to \mathbf{B} - {\rm d}_{\mathbf{A}}\boldsymbol{\eta}. \label{secondGauge}
\ee
This latter set of local invariances is so constraining that BF theory has no local dynamics---the theory is purely topological. Of course, gravity is not a topological theory, and in the BF coupled gravity model, this second set of invariances will be explicitly broken.

\subsection{BF-Coupled Gravity}

As proposed in \cite{Alexander:2018tyf, alexander2020chiral}, we consider a theory of gravity in four spacetime dimensions obtained by replacing the ``cosmological constant" $\mu$ in the BF action with the usual Lagrangian of Einstein gravity written in terms of a composite metric $\hat{g}_{\mu\nu}$, which we define below. The action of the theory is\footnote{The sign for $\bar{\Lambda}$ differs here from that used in  \cite{Alexander:2018tyf, alexander2020chiral} for reasons that will become clear later, namely, $\Lambda(r \to \infty) = \bar\Lambda$, rather than $\Lambda(r \to \infty) = -\bar\Lambda$.}
\begin{equation}
S_0 =\int_M {\rm tr}(\mathbf{B}\wedge \mathbf{F})+\biggl[\frac{1}{2\kappa}R(\hat{g})-\frac{\bar{\Lambda}}{\kappa}+\mathcal{L}_M\biggr] {\rm tr}(\mathbf{B}\wedge\mathbf{B}) \label{action}
\end{equation}
where $\kappa = 8\pi G$ is Einstein's constant, and we impose the requirement that $\mathbf{B}$ is \emph{non-degenerate}---i.e. ${\rm{tr}(\mathbf{B}\wedge\mathbf{B}) \neq 0}$. Now ${\rm tr}(\mathbf{B}\wedge\mathbf{B})$ is a volume form on $M$, which determines a density $\omega$, represented in a given coordinate system by 
\begin{equation}
{\rm tr}(\mathbf{B}\wedge\mathbf{B}) = \frac{\sqrt{\omega}}{4!}\epsilon_{\mu\nu\rho\sigma}dx^\mu\!\wedge dx^\nu\!\wedge dx^\rho\!\wedge dx^\sigma. \label{omegaDef}
\end{equation}
The composite metric $\hat{g}_{\mu\nu}$ is then defined as
\begin{equation}
\hat{g}_{\mu\nu} = \left(\frac{\omega}{g}\right)^{\!1/4}\!g_{\mu\nu}, \label{vol}
\end{equation}
where $g_{\mu\nu}$ is an arbitrary metric, which we take to be one of the fundamental fields upon which the action \eqref{action} depends. Note that, by construction, $\hat{g}_{\mu\nu}$ is invariant under Weyl transformations of the bare metric 
\be
g_{\mu\nu} \to \Omega^2g_{\mu\nu},
\ee
and thus, so is the action $S_0$, since it is a functional of $\hat{g}_{\mu\nu}$. Varying \eqref{action} with respect to $\mathbf{B}$, $\mathbf{A}$ and $g_{\mu\nu}$ yields
\begin{gather}
\mathbf{F}+\frac{1}{2\kappa}[R(\hat{g})- 4\bar\Lambda + \kappa T]\mathbf{B}=0 \label{EOM1}\\
{\rm d}_{\mathbf{A}} \mathbf{B}=0 \label{EOM2}\\
R_{\mu\nu}(\hat{g})-\frac{1}{4}R(\hat{g})\hat{g}_{\mu\nu}=\kappa\biggl(T_{\mu\nu}-\frac{1}{4}T\hat{g}_{\mu\nu}\biggr), \label{TFEE}
\end{gather}
where 
\begin{equation}
\label{defTab}
T_{\mu\nu}=-\frac{2}{\sqrt{|\omega|}}\frac{\delta\mathcal{S}_{\rm M}(\hat{g},\bm{\Phi})}{\delta\hat{g}^{\mu\nu}}=-2\frac{\partial\mathcal{L}_{\rm M}}{\partial \hat{g}^{\mu\nu}}+\mathcal{L}_{\rm M}\hat{g}_{\mu\nu}.
\end{equation}
For details, see \cite{Alexander:2018tyf}. Equation \eqref{TFEE} is the \emph{traceless} Einstein equation. The Ricci scalar is determined by the remaining equations as follows. Taking the covariant exterior derivative of \eqref{EOM1}, and using equation \eqref{EOM2}, one obtains
\be
\label{dEOM}
{\rm d}(R(\hat{g})+\kappa T) = 0,
\ee
which can be integrated to give
\be
\label{RTLambda}
R(\hat{g})+\kappa T = 4 \Lambda,
\ee
where $\Lambda$ is an integration constant and the factor of four is chosen so that when \eqref{RTLambda} is inserted into \eqref{TFEE}, one obtains the full Einstein equation, with the trace part restored, and with $\Lambda$ appearing as the physical cosmological constant. It is worth emphasizing that $\bar\Lambda$, which appears in the Lagrangian where the cosmological constant ordinarily would, and which is subject to loop corrections, does not play the role of a cosmological constant at the level of the equations of motion---it is simply a coupling constant that appears only in the BF sector of the theory. As noted in \cite{Alexander:2018tyf, alexander2020chiral}, this theory can provide an explanation for why the cosmological constant is not renormalized to a large value by quantum corrections to the vacuum energy from quantum fields---the coupling constant $\bar\Lambda$ that receives quantum corrections from the vacuum energy of quantum fields does not gravitate, and the constant $\Lambda$ that gravitates does not receive quantum corrections. This mechanism is similar in certain respects to the ``sequestering" approach proposed by Kaloper and Padilla \cite{kaloper2014sequestering}. 

\section{BF-coupled gravity with a gauge kinetic term}
We now consider a minimal extension of the BF-coupled gravity model presented in the previous section by including a kinetic term for the gauge field $\mathbf{A}$ 
\begin{eqnarray}
S = S_0  - \frac{1}{2g^2}\int_M{\rm tr}\left(\mathbf{F}\wedge\hat{\star}\,\mathbf{F}\right), \label{NewAction}
\end{eqnarray}
where $g$ is the gauge coupling, and $\hat{\star}$ is the Hodge star operator associated with the composite metric $\hat{g}_{\mu\nu}$. Viewing this model as a modification of BF theory, it seems natural upon first consideration \emph{not} to include a gauge kinetic term, since no such term can be constructed without introducing a metric and spoiling the topological nature of the theory. However, once one couples BF theory to gravity, local degrees of freedom are already present, a dynamical spacetime metric is available, and one can naturally include a gauge and diffeomorphism invariant kinetic term for the gauge field. We will see that including this term naturally sources fluctuations of the effective cosmological constant $\Lambda$ defined in \eqref{RTLambda}.

Varying \eqref{NewAction} with respect to $\mathbf{B}$, $\mathbf{A}$ and $g_{\mu\nu}$, one finds
\begin{gather}
\mathbf{F}+\tfrac{1}{2\kappa}[R(\hat{g}) - 4\bar\Lambda + \kappa T]\mathbf{B}=0 \label{EOM1.2}\\
{\rm d}_{\mathbf{A}} \mathbf{B}= \tfrac{1}{g^2}{\rm d}_{\mathbf{A}} \hat{\star}\,\mathbf{F} \label{EOM2.2}\\
R_{\mu\nu}(\hat{g})-\tfrac{1}{4}R(\hat{g})\hat{g}_{\mu\nu}=\kappa\left(T_{\mu\nu}-\tfrac{1}{4}T\hat{g}_{\mu\nu}\right), \label{TFEE2}
\end{gather}
where $T_{\mu\nu}$ is the stress tensor associated with the physical inverse metric $\hat{g}^{\mu\nu}$
\begin{equation}
T_{\mu\nu}=-\frac{2}{\sqrt{|\omega|}}\frac{\delta\mathcal{S}_{\rm M}(\hat{g},\bm{\Phi})}{\delta\hat{g}^{\mu\nu}}=-2\frac{\partial\mathcal{L}_{\rm M}}{\partial \hat{g}^{\mu\nu}}+\mathcal{L}_{\rm M}\hat{g}_{\mu\nu}, \label{stress}
\end{equation}
and $T:= \hat{g}^{\mu\nu}T_{\mu\nu}$ is its trace. The presence of the gauge kinetic term alters equations \eqref{EOM1} and \eqref{TFEE} only in that the stress tensor now includes contributions from the gauge field as well as other matter fields, but their form is otherwise unchanged. Meanwhile, \eqref{EOM2.2} reproduces \eqref{EOM2} of the original model in the limit $g \to \infty$. Defining $\Lambda = \tfrac{1}{4}(R(\hat{g})+\kappa T)$, combining equations \eqref{EOM1.2} and \eqref{EOM2.2}, and making use of the Bianchi identity ${\rm d}_{\mathbf{A}} \mathbf{F} = 0$, we obtain a simple equation for the field strength two-form~$\mathbf{F}$:
\be
\frac{{\rm d}\Lambda\wedge\mathbf{F}}{(\Lambda - \bar{\Lambda})^2} = \tfrac{2}{g^2 \kappa}{\rm d}_{\mathbf{A}} \hat{\star}\,\mathbf{F}. \label{F-eq}
\ee
In the following sections, we will study the spherically symmetric black hole solutions and homogeneous, isotropic cosmological models predicted by this model. In these simple examples, we will see that the ``cosmological constant" $\Lambda$ will be promoted to a function of $r$ in the black hole case, and a function of $t$ in the cosmological case.

\subsection{Spherically symmetric solutions}

Let us consider the model described by the action \eqref{NewAction} with an Abelian gauge group, say $U(1)$. While we restrict to the Abelian case for simplicity, it should be noted that the solutions we find in this section can be trivially extended to solutions of the theory with a non-Abelian gauge group, simply by multiplying $\mathbf{A}$, $\mathbf{B}$ and $\mathbf{F}$ by the same group generator. For instance, if we wanted to embed the Abelian solution into the theory with gauge group $SU(2)$, we could simply multiply each of these fields by $\sigma_1$. Later, we will consider a cosmological model which can only be realized in a non-Abelian gauge group---in particular, we focus on the case $G = SU(2)$---but it should be stressed that the solutions in this section may be regarded as solutions of the \emph{same theory} after performing the trivial embedding we have just described. 

We would like to find stationary, spherically symmetric solutions with radially-varying ``cosmological constant" $\Lambda = \Lambda(r)$. As a first step, let us consider the Abelian version of equation \eqref{F-eq}
\be
\frac{{\rm d}\Lambda\wedge\mathbf{F}}{(\Lambda - \bar{\Lambda})^2} = \tfrac{2}{g^2 \kappa}{\rm d}\, \hat{\star}\,\mathbf{F}. \label{abel-F-eq}
\ee
In order to evaluate the right hand side of \eqref{abel-F-eq}, we need to specify the physical metric $\hat{g}_{\mu\nu}$. For this, we consider the general stationary, spherically symmetric line element
\be
ds^2 = -\alpha(r)^2 dt^2 + \beta(r)^2 dr^2 + r^2d\Omega^2, \label{metric}
\ee
where $ds^2 = \hat{g}_{\mu\nu}dx^\mu dx^\nu $, and  $d\Omega^2 = d\theta^2 + \sin^2\theta \, d\phi^2$ is the metric on the unit two-sphere. In order to reproduce the volume form associated with \eqref{metric}, we can consider a $\BB$ field of the form
\be
\BB = \frac{1}{\sqrt{2}}\Big(\alpha\beta r^2 (\Lambda - \bar{\Lambda})dt \wedge dr + \frac{\sin\theta}{(\Lambda - \bar\Lambda)} d\theta \wedge d\phi \Big). \label{B}
\ee
Making use of \eqref{EOM1.2}, this gives
\bea
\BF &=& -\frac{2}{\kappa}(\Lambda - \bar\Lambda)\BB \nn \\
&=& -\frac{\sqrt{2}}{\kappa}\left[ \alpha\beta r^2 (\Lambda - \bar\Lambda)^2 dt\wedge dr + \sin\theta \, d\theta\wedge d\phi \right], \nn \\ \label{F}
\eea
and it can be checked that ${\rm d}\BF = 0$, since $\Lambda$ depends only on $r$.
Using \eqref{metric}, we can compute the Hodge star of \eqref{F}
\be
\hat{\star}\,\BF = -\frac{\sqrt{2}}{\kappa}\left[\frac{\alpha\beta}{r^2}dt\wedge dr - (\Lambda - \bar\Lambda)^2 r^4 \sin\theta \, d\theta\wedge d\phi \right].
\ee
Now, from \eqref{F-eq} we have 
\be
 \frac{d}{dr}\Big(\frac{1}{\Lambda - \bar\Lambda}\Big) = \frac{2}{ g^2\kappa}\frac{d}{dr}\left[(\Lambda - \bar\Lambda)^2 r^4\right],
\ee
which can be integrated immediately to give
\be
\frac{1}{\Lambda - \bar\Lambda} = \frac{2}{g^2\kappa}(\Lambda - \bar\Lambda)^2 r^4 + \frac{1}{\Lambda_0 - \bar\Lambda},
\ee
where $\Lambda_0$ is an integration constant. This can be rearranged to give a cubic equation for $\Lambda - \bar\Lambda$
\be
\frac{2}{g^2\kappa}(\Lambda - \bar\Lambda)^3 r^4 + \frac{\Lambda - \bar\Lambda}{\Lambda_0 - \bar\Lambda} - 1 = 0, \label{cubic}
\ee
which fixes $\Lambda - \bar\Lambda$ as a function of $r$. Since \eqref{cubic} is cubic, it has three solutions, but only one of them is real for all $r > 0$. There is a fixed, positive value $r = r_{\mathrm{max}}$ for which the remaining two solutions degenerate to a real, double root. For $0 < r < r_{\mathrm{max}}$, there are three distinct real roots, and for $r > r_{\mathrm{max}}$ there is a single real root and a pair of complex conjugate roots. In what follows, we will focus primarily on the solution that is defined for all $r > 0$, and comment briefly on the other solutions in the discussion section. The solution that is real for all $r > 0$ has the explicit form 

\be
    \Lambda - \bar{\Lambda} = \frac{Q(X)}{\sqrt[3]{2}~3^{2/3}X^4} - \frac{\sqrt[3]{\frac{2}{3}}\mathcal{B}}{Q(X)},
    \label{LpLb}
\ee
where
\be
Q(X) := \sqrt[3]{9 X^8 + \sqrt{3}\sqrt{27 X^{16} + 4\mathcal{B}^3 X^{12}}} \, ,
\ee
and we have defined $X^4 = 2(g^2 \kappa)^{-1} r^4$, and $\mathcal{B} = (\Lambda_0 - \bar{\Lambda})^{-1}$. When $\Lambda_0 - \bar\Lambda > 0$, we have $\Lambda(0) = \Lambda_0$. When $\Lambda_0 - \bar\Lambda < 0$, $\Lambda$ diverges as $r$ goes to zero. When $\Lambda_0 - \bar\Lambda = 0$, we have $\Lambda = \Lambda_0 = \bar\Lambda$, which can be seen by taking the appropriate limit of \eqref{cubic}, and the solution reduces to the (A)dS--Schwarzshild metric.

Next we turn to the trace-free Einstein equation \eqref{TFEE}. The stress tensor associated with $\BF$ is  
\be
T_{\mu\nu} = \frac{1}{g^2}\Big( F^\rho{}_{\!\mu} F_{\rho\nu} - \frac{1}{4}F^{\rho\sigma}F_{\rho\sigma}\hat{g}_{\mu\nu} \Big). \label{Tgauge}
\ee
Inserting \eqref{F} into \eqref{Tgauge} gives
\be
T_{\mu\nu}dx^\mu dx^\nu = \frac{1}{g^2\kappa^2r^4}[1 + r^8(\Lambda - \bar\Lambda)^4](\alpha^2 dt^2 - \beta^2 dr^2 + r^2d\Omega^2), \label{Tgauge2}
\ee
from which it follows that $T = 0.$ Meanwhile, the left-hand side of equation \eqref{TFEE} becomes
\begin{widetext}
\bea
\Big(R_{\mu\nu} - \frac{1}{4}R \hat{g}_{\mu\nu}\Big)dx^\mu dx^\nu = \frac{1}{2r^2\alpha\beta^3}(\alpha''\beta r^2 + \alpha\beta^3 - \alpha'\beta' r^2 - \alpha\beta)(\alpha^2 dt^2 - \beta^2 dr^2 + r^2 d\Omega^2) + \frac{\alpha}{r\alpha\beta^3}(\alpha\beta)'(\alpha^2 dt^2 + \beta^2 dr^2). \nn \\
\eea
\end{widetext}
Both the $\theta\theta$ and $\phi\phi$ components of \eqref{TFEE} lead to 
\be
\frac{1}{2r^2\alpha\beta^3}(\alpha''\beta r^2 + \alpha\beta^3 - \alpha'\beta' r^2 - \alpha\beta) = \frac{1}{g^2\kappa r^4}[1 + r^8(\Lambda - \bar\Lambda)^4]. \label{angles}
\ee
Inserting \eqref{angles} into either the $tt$ or $rr$ equation then leads to 
\be
(\alpha\beta)' = 0,
\ee
which can be immediately integrated to give 
\be
\alpha\beta = C,
\ee
and by rescaling the time coordinate, we can always set $C = 1$, so that 
\be
\beta = \frac{1}{\alpha}. \label{killBeta}
\ee
Putting \eqref{killBeta} back into \eqref{angles}  we obtain the following equation for $\alpha$
\be
(\alpha\alpha')' + \frac{1-\alpha^2}{r^2} = \frac{2}{g^2\kappa r^4}[1 + r^8(\Lambda - \bar\Lambda)^4]. \label{alphaOnly}
\ee
The above equation is a linear, inhomogeneous equation for $\alpha^2$ whose homogeneous solution reproduces the (A)dS-Schwarzschild metric. Let us define 
\be
f(r) := \frac{2}{g^2\kappa r^4}[1 + r^8(\Lambda - \bar\Lambda)^4],
\ee
so that the equation for $A := \alpha^2$ now reads
\be
\frac{1}{2}A'' + \frac{1-A}{r^2} = f(r). \label{generalInhomogeneousEq}
\ee
This last equation has a compact general solution for any function $f(r)$, given by 
\be
A = 1 - \frac{\Lambda_*}{3}r^2 - \frac{2M}{r} + \frac{2r^2}{3}\!\int_{r_0}^r \frac{f(s)}{s}ds - \frac{2}{3r}\!\int_{r_0}^r f(s)s^2 ds, \label{generalInhomogeneousSolution}
\ee
where $r_0$ is an arbitrary valid position where we define $\Lambda_*$ and $M$, the integration constants. Note that $r_0$ does not need to be the same for both integrals.

Since the metric function $A$ is the solution of a second order linear equation, the lower endpoint of the integrals is not independent of the integration constants $\Lambda_*$ and $M$. It would therefore be advantageous to fix the lower endpoint of integration to a convenient value.

First, it will be convenient to perform the integrals over the term in $f(r)$ that is proportional to $r^{-4}$. This leads to 
\be
A = 1 - \frac{2M}{r} - \frac{\Lambda_*}{3}r^2 + \left(\frac{1}{g^2\kappa}\right)\frac{1}{r^2} + I_1(r) + I_2(r), \label{Reissner-Nordstrom-term}
\ee
where
\bea
I_1 &=& \frac{4 r^2}{3 g^2 \kappa}\int_{r_0}^rs^3(\Lambda(s) - \bar\Lambda)^4 ds, \label{eq:integral_1} \\
I_2 &=& -\frac{4}{3g^2\kappa r}\int_{r_0}^r s^6 (\Lambda(s) - \bar\Lambda)^4 ds, \label{eq:integral_2}
\eea
and we have absorbed $r_0$-dependent terms coming from the integration into a redefinition of the integration constants $\Lambda_*$ and $M$.\footnote{Actually, one may choose the lower endpoint of each integral independently, in which case choosing the lower endpoint at infinity produces no shift in $\Lambda_*$ and $M$.} Furthermore, the integrals in $I_1$ and $I_2$ can be expressed as
\begin{align}
\int_{r_0}^r s^3 & (\Lambda - \bar\Lambda)^4 ds = -\frac{g^2 \kappa}{4}(\Lambda - \bar{\Lambda}) -\frac{s^4}{4} (\Lambda -\bar{\Lambda})^4 \bigg|_{s=r_0}^r \label{eq:I_1_exact} \\
\int_{r_0}^r s^6 & (\Lambda - \bar\Lambda)^4 ds = \frac{1}{5} s^7 (\Lambda - \bar\Lambda)^{4} \nonumber \\
& \times \left[ 3 - 16 \frac{\Lambda - \bar{\Lambda}}{\Lambda_0 - \bar{\Lambda}} \; {}_2 F_1 \! \left(1,\frac{3}{2};\frac{3}{4}; \frac{\Lambda - \bar{\Lambda}}{\Lambda_0 - \bar{\Lambda}} \right) \right] \bigg|_{s=r_0}^r \label{eq:I_2_exact}
\end{align}
where ${}_2 F_1(a,b\,;c\,;z)$ is the hypergeometric function (see \AppRef{ap:integration}).
\footnote{As with the $r^{-4}$ term in equation \eqref{Reissner-Nordstrom-term}, we can always shift $r_0$ and adjust $\Lambda_*$ and $M$ to compensate. For equation \eqref{eq:I_1_exact}, $r_0 \to \infty$ once again leads to no adjustments in the parameters. For equation \eqref{eq:I_1_exact}, this point corresponds to the position where $16x \, {}_2 \! F_1 \! (1,3/2;3/4;x)=3$ is satisfied, and $x=(\Lambda-\bar{\Lambda})/(\Lambda_0-\bar{\Lambda})$.}

Now we proceed to fix $\Lambda_*$ in a manner that is consistent with the $r \to \infty$ limit. For this, we compute the Ricci scalar associated with the metric
\be
ds^2 = -A(r) dt^2 + \frac{1}{A(r)} dr^2 + r^2d\Omega^2. \label{metric2}
\ee
For any metric of the form \eqref{metric2}, one has
\be
-4\Lambda = -R = A'' + \frac{4}{r}A' + \frac{2}{r^2}(A - 1),
\ee
and with $A(r)$ given by equation \eqref{Reissner-Nordstrom-term}, this becomes
\be
\Lambda = \Lambda_* - \frac{4}{g^2\kappa}\int_{r_0}^r s^3(\Lambda - \bar\Lambda)^4 ds - \frac{r^4}{g^2\kappa}(\Lambda - \bar\Lambda)^4. \label{consistency}
\ee
If we set $r \to r_0$ and then take the limit as $r_0 \to \infty$, we find that $\Lambda_*= \bar{\Lambda}$, since $\Lambda - \bar{\Lambda}$, $r^4(\Lambda - \bar{\Lambda})^4$, and the integrals all go to zero in these limits. We thus have
\be
\frac{4}{g^2\kappa}\int_{r}^\infty s^3(\Lambda - \bar\Lambda)^4 ds = (\Lambda - \bar{\Lambda}) + \frac{r^4}{g^2\kappa}(\Lambda - \bar\Lambda)^4 \label{eq:int1_explicite}
\ee
in complete agreement with equation \eqref{eq:I_1_exact} with $r_0 \to \infty$. For $\Lambda_0 - \bar{\Lambda} > 0$, the integral also has the property
\be
\frac{4}{g^2\kappa}\int_0^\infty s^3(\Lambda - \bar\Lambda)^4ds = \Lambda_0 - \bar\Lambda. \label{exactInt}
\ee

It is instructive to examine the integrals in equations \eqref{eq:I_1_exact} and \eqref{eq:I_2_exact}. Up to leading order, and ignoring any terms that can be shifted into $\Lambda_*$ or $M$, the integral contributions are:
\begin{align}
    \intertext{$\mathrm{for} \quad r \ll r_c \quad \mathrm{and} \quad \Lambda_0-\bar{\Lambda}>0$}
    I_1 + I_2 \approx & \, \frac{(\Lambda_0-\bar{\Lambda})^2}{7g^2 \kappa}r^6, \\
    \intertext{$\mathrm{for} \quad r \ll r_c \quad \mathrm{and} \quad \Lambda_0-\bar{\Lambda}<0$}
    I_1 + I_2 \approx & \, \frac{g^2\kappa}{4(\Lambda_0-\bar{\Lambda})^2}r^{-2}, \label{eq:asymptotic_small_r_neg_lambda} \\
    \intertext{$\mathrm{for} \quad r \gg r_c$}
    I_1 + I_2 \approx & \, -\frac{9}{10} \left( \frac{g^2 \kappa}{2} \right)^{1/3} r^{2/3},
\end{align}
where $r_c = |g^2 \kappa/2 (\Lambda_0 - \bar{\Lambda})^3|^{1/4}$, which corresponds to the location where the two limiting solutions of $\Lambda - \bar{\Lambda}$ intersect. For $r \ll r_c$ and $\Lambda_0-\bar{\Lambda}>0$, the integrals are subdominant to the $\Lambda_*$, $M$, and $r^{-2}$ terms and mainly constitute a rapid transition between the near and far behaviors. For $r \gg r_c$, the integrals are once again subdominant to the $\bar{\Lambda}r^2$ term, and thus only dominate if $\bar{\Lambda}=0$. However, if $\bar{\Lambda}=0$, this $r^{2/3}$ dependence is a novel property of this theory, and the asymptotic structure of these solutions warrants further investigation.

Finally, for the $r \ll r_c$ and $\Lambda_0-\bar{\Lambda}<0$ case, the integral behavior dominates, producing an $r^{-2}$ term which combines with the one already present. Interestingly, this term is exactly the term proportional to the total electric charge of a Reissner-Nordstr\"{o}m black hole. However, in this model the divergence of the field strength is sourced by $d\Lambda$ rather than an external current, so there is no notion of total charge, and the strength of the $r^{-2}$ term is controlled purely by the combination $\Lambda_0 -\bar{\Lambda}$ and the couplings $g$ and $\kappa$. This is an interesting feature---for the case $\Lambda_0 - \bar\Lambda > 0$, this model predicts black holes that display a certain universal behavior, as though each had an equal charge of unit magnitude, much like a fundamental particle. For the case $\Lambda_0 - \bar\Lambda <0$ the situation is a bit subtler, and the effective charge depends on the couplings as well as the particular value of $\Lambda_0 - \bar\Lambda$, which can be seen from equation \eqref{eq:asymptotic_small_r_neg_lambda}.
The analogy between black holes and fundamental particles has been remarked upon since the discovery of the no-hair theorems \cite{Carter:1971zc, heusler1996black} and was recently explored in a somewhat different context \cite{arkani2020kerr}, but to our knowledge, the  effective ``charge quantization" we present here is a novel feature.  Of course, it should be acknowledged that we assumed stationarity and spherical symmetry in obtaining this solution, so it is not clear whether this universal behavior is a robust prediction for black holes in this model, or merely an artifact of enhanced symmetry. It would be interesting to see whether this feature persists in black hole mergers, or whether there is an additive property to the coefficient of the $r^{-2}$ term in such cases, as one would expect from charge conservation.

\begin{figure}[ht]
     \includegraphics[width=0.9\columnwidth]{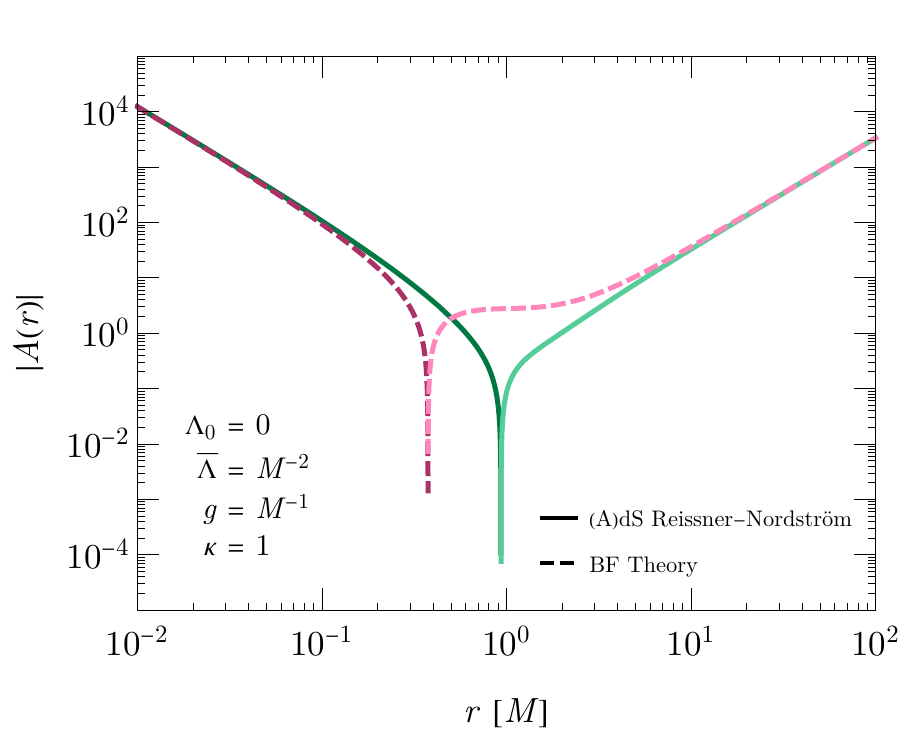}
    \caption{The BF spherically symmetric metric function compared with its corresponding (A)dS Reissner-Nordstr\"{o}m counterpart. The charge is set to match the small $r$ asymptotic behavior ($Q^2 = 1/g^2 \kappa$ or $1/g^2 \kappa + g^2 \kappa/4(\Lambda_0 - \bar{\Lambda})^2$ for positive and negative $\Lambda_0 - \bar{\Lambda}$ respectively).
    }
    \label{fig:spherical_metric_comp}
\end{figure}

In \figref{fig:spherical_metric_comp}, we show an example of the spherically symmetric BF theory metric function, $A(r)$. For comparison, we also plot the corresponding (A)dS Reissner-Nordstr\"{o}m black hole counterpart
\begin{equation}
    A(r) = 1 - \frac{2M}{r} - \frac{\bar{\Lambda}}{3}r^2 + \frac{Q^2}{r^2}
    \label{eq:reissener_nordstrom}
\end{equation}
where $Q$ is the charge of the black hole. The main qualitative differences are a change to the metric function near the event horizon as well as a change in the location of the horizon. As noted previously, the large $r$ behavior in the two cases will differ dramatically for $\bar{\Lambda}=0$ as \eqref{eq:reissener_nordstrom} will be asymptotically flat while \eqref{Reissner-Nordstrom-term} is not.

Calculating the Kretschmann invariant for a metric of the form \eqref{metric2} leads to a simple expression
\begin{equation}
R_{\mu \nu \rho \sigma}R^{\mu \nu \rho \sigma} = (A'')^2 + \frac{4}{r^2}(A')^2 + \frac{4}{r^4}(A-1)^2.
\end{equation}
Using the metric described here, the Kretschmann invariant detects no physical singularities except at $r=0$, as expected.
The $r \to 0$ behavior goes as $r^{-8}$ coming from the $r^{-2}$ portions of the metric as in the Reissner-Nordstr\"{o}m solution. The $r \to \infty$ behavior goes as $8\bar{\Lambda}^2/3+\mathcal{O}(r^{-4/3})$, exhibiting (A)dS like behaviors with a higher order correction.

\subsection{A Homogenous, Isotropic Cosmological Model}
We now turn to a homogeneous, isotropic cosmological model, where we assume for simplicity that the stress tensor is sourced purely by the gauge field. We leave a detailed analysis including contributions to the stress tensor from other sources for future work.

One might hope to find a simple, homogeneous, isotropic cosmological model including the gauge kinetic term for an Abelian $U(1)$ or $\mathbb{R}_+$ gauge group. However, because of the restriction that tr$(\bf{B}\wedge\bf{B}) \neq 0$, and hence tr${(\bf{F}\wedge\bf{F}) \neq 0}$, it is not possible to preserve isotropy in the Abelian case. On the other hand, homogeneity and isotropy can be preserved by a simple ansatz for a gauge field $\bf{A}$ taking values in the Lie algebra of $SU(2)$:
\be
\mathbf{A} = \alpha(t)\left(\tau_1 dx + \tau_2 dy + \tau_3 dz\right)
\ee
where $\tau_i$ are the group generators, given by $\tau_i = -\tfrac{i}{2}\sigma_i$, and $\sigma_i$ are Pauli matrices. This ansatz was previously employed in the context of chromonatural inflation \cite{adshead2013perturbations, maleknejad2016axion} and in a toy model of dark energy where the energy density of the universe today is dominated by a pseudo-scalar axion which couples to the Pontryagin density of a gauge field \cite{alexander2016tracking}. The corresponding field strength two-form is
\bea
{\rm \hspace{-.75cm}}  
\mathbf{F} := {\rm d}\mathbf{A} + \mathbf{A}\wedge\mathbf{A}  &=& \left(\dot{\alpha} dt\wedge dx + \alpha^2 dy\wedge dz\right)\!\tau_1 \nn \\
&~+~& \left(\dot{\alpha} dt\wedge dy + \alpha^2 dz\wedge dx\right)\!\tau_2 \nn \\
 &~+~& \left(\dot{\alpha} dt\wedge dz + \alpha^2 dx\wedge dy\right)\!\tau_3. \label{Fsu2}
\eea
For later use, we note that
\be
\textrm{tr}(\mathbf{F}\wedge\mathbf{F}) = -3\dot\alpha\alpha^2d^4 x.
\ee 
If we assume the usual spatially flat FRW form for the spacetime metric
\be
ds^2 = \hat{g}_{\mu\nu}dx^\mu dx^\nu = -dt^2 + a(t)^2(dx^2 + dy^2 + dz^2), \label{frw}
\ee
we can also compute the Hodge star of the field strength
\bea
\hat{\star}\,\mathbf{F} &=& \left( \frac{\alpha^2}{a}dt\wedge dx - a\dot\alpha dy\wedge dz \right)\!\tau_1 \nonumber \\
&~~~~~+& \left( \frac{\alpha^2}{a}dt\wedge dy - a\dot\alpha dz\wedge dx \right)\!\tau_2 \nonumber \\
 &~~~~~+& \left(\frac{\alpha^2}{a}dt\wedge dz - a\dot\alpha dx\wedge dy\right)\!\tau_3.
\eea

We now turn to equation \eqref{F-eq}, which we rewrite in the form
\be
{\rm d}u \wedge \mathbf{F} = {\rm d}_{\mathbf{A}}\hat{\star}\,\mathbf{F}, \label{nice}
\ee
where we have defined 
\be{}
u = -\frac{g^2 \kappa}{2(\Lambda - \bar{\Lambda})}.
\label{eq:u_cosmo}
\ee

Recall that for any adjoint-valued two-form $\boldsymbol{\Omega}$, of which $\hat{\star} \, \mathbf{F}$ is an example, we have
\be
{\rm d}_{\mathbf{A}}\boldsymbol{\Omega} = {\rm d}\boldsymbol{\Omega} + \mathbf{A}\wedge \boldsymbol{\Omega} - \boldsymbol{\Omega}\wedge \mathbf{A},
\ee
from which we can compute the right-hand side of \eqref{nice}:
\begin{widetext}
\be
{\rm d}_{\mathbf{A}}\hat{\star}\,\mathbf{F} = -\left[ 2\frac{\alpha^3}{a} + \frac{d}{dt}(a\dot\alpha) \right](dt\wedge dy\wedge dz \,\tau_1 + dt\wedge dz\wedge dx\, \tau_2 + dt\wedge dx\wedge dy \,\tau_3). \label{RHS1}
\ee
Meanwhile, 
\bea
{\rm d}u \wedge \mathbf{F} &=& \left[\dot\alpha(u_y dt\wedge dx\wedge dy - u_z dt\wedge dz\wedge dx) + \alpha^2(\dot{u} dt\wedge dy\wedge dz + u_x dx\wedge dy\wedge dz)\right]\tau_1 \nonumber \\
&~~~~~+& \left[\dot\alpha(u_z dt\wedge dy\wedge dz - u_x dt\wedge dx\wedge dy) + \alpha^2(\dot{u} dt\wedge dz\wedge dx + u_y dx\wedge dy\wedge dz)\right]\tau_2 \nonumber \\
&~~~~~+& \left[\dot\alpha(u_x dt\wedge dz\wedge dx - u_y dt\wedge dy\wedge dz) + \alpha^2(\dot{u} dt\wedge dx\wedge dy + u_z dx\wedge dy\wedge dz)\right]\tau_3. \nonumber \\ \label{LHS1}
\eea
\end{widetext}
Putting \eqref{RHS1} and \eqref{LHS1} into \eqref{nice}, we find that
\begin{gather}
\dot{u} = -2\frac{\alpha}{a} - \frac{1}{\alpha^2}\frac{d}{dt}(a\dot\alpha), \qquad
u_x = u_y = u_z = 0, \label{udot}
\end{gather}
which implies that $\Lambda$ depends only on time, which confirms that our ansatz is compatible with homogeneity.
Furthermore, we require that 
\be
\textrm{tr}(\mathbf{B}\wedge\mathbf{B}) = \sqrt{-\hat{g}}\,d^4 x = a^3\,d^4 x, \label{vol1}
\ee
while equation \eqref{EOM1} implies that 
\be
\textrm{tr}(\mathbf{B}\wedge\mathbf{B}) = \frac{\kappa^2}{4(\Lambda - \bar{\Lambda})^2}\textrm{tr}(\mathbf{F}\wedge\mathbf{F}) = \frac{-3\kappa^2\dot\alpha \alpha^2}{4(\Lambda - \bar{\Lambda})^2}\,d^4x \label{vol2}.
\ee
Combining equations \eqref{vol1} and \eqref{vol2} then leads to 
\be
 \frac{-3\kappa^2\dot\alpha \alpha^2}{4(\Lambda - \bar{\Lambda})^2} = a^3.
 \label{eq:lambda_lambdabar_a_relation}
\ee
Rewriting the left-hand side in terms of $u$ and solving for $u^2$, we have
\be
u^2 = \frac{-a^3g^4}{3\dot\alpha\alpha^2}. \label{uSquared}
\ee
Now we compute the stress tensor and investigate the model numerically.

The stress tensor for the gauge field is
\begin{equation}
    T_{\mu\nu} = \frac{1}{g^2} {\rm tr}\left( F^\rho_{~\mu} F_{\rho \nu} - \frac{1}{4} g_{\mu\nu} F^{\rho \sigma}F_{\rho \sigma} \right). \label{stress-su2}
\end{equation}
Now plugging \eqref{Fsu2} into \eqref{stress-su2} we have,
\begin{align}
    T_{\mu \nu}dx^\mu dx^\nu & = - \frac{1}{4g^2} \left( \frac{\dot{\alpha}^2}{a^2} + \frac{\alpha^4}{a^4} \right) \nonumber \\
    & \times \left[~3 dt^2 + a^2 dx^2 + a^2 dy^2 + a^2 dz^2 ~ \right],
    \label{tmunu-su2}
\end{align}
from which it follows that $T = 0$. One can then compute the Friedmann equation after plugging \eqref{tmunu-su2} and \eqref{frw} into \eqref{TFEE},
\begin{equation}
    a \ddot{a} - \dot{a}^2 = \frac{\kappa}{2g^2}\left(\dot{\alpha}^2 + \frac{\alpha^4}{a^2}\right).
    \label{friedmann}
\end{equation}
Clearly, $\ddot{a}>0$. Thus the model permits a single bounce consistent with a $\Lambda$ dominated universe. Note that the trace-free property of the Einstein equation reduces the number of Friedmann equations from two to one. Combining \eqref{udot} and \eqref{uSquared} yields
\begin{equation}
    \left( \frac{g^4 a^3}{3 \dot{\alpha}^2} + 2 u a \right) \ddot{\alpha} = 2 u^2 \alpha \dot{\alpha} + \frac{g^4 a^2\dot{a}}{\dot{\alpha}} - \frac{4 u \alpha^3}{a} - 2 u \dot{a} \dot{\alpha},
    \label{eq:alphadotdot}
\end{equation}
and together with \eqref{friedmann}, we get two coupled differential equations that fully describe the dynamics of the system up to the sign of $u$.

Furthermore, using the definition of $\Lambda$, the trace-free property of the Yang-Mills stress tensor, and the FRW form of the metric, we obtain
\begin{equation}
    \Lambda = R/4 = \frac{3}{2a^2} \left( \dot{a}^2 + a \ddot{a} \right)
\end{equation}
which, using \eqref{eq:lambda_lambdabar_a_relation} and \eqref{friedmann}, leads to
\begin{equation}
    \Lambda = \frac{3}{2a^2} \left( 2\dot{a}^2 + \frac{\kappa}{2g^2}\left(\dot{\alpha}^2 + \frac{\alpha^4}{a^2}\right) \right)
    \label{eq:Lambda_cosmo}
\end{equation}
and
\begin{equation}
{\rm \hspace{-0.05cm}}    \bar{\Lambda} = \mp \sqrt{\frac{-3 \kappa^2 \dot{\alpha} \alpha^2}{4 a^3}} + \frac{3}{2a^2} \left( 2\dot{a}^2 + \frac{\kappa}{2g^2}\left(\dot{\alpha}^2 + \frac{\alpha^4}{a^2}\right) \right)
    \label{eq:Lambdabar_cosmo}
\end{equation}
where the $\mp$ symbol is understood to indicate the negative sign when $u > 0$ and the positive sign when $u < 0$. 
Thus, allowing the calculation of the sign of $u$ at all times using \eqref{eq:u_cosmo}, \eqref{eq:Lambda_cosmo}, and \eqref{eq:Lambdabar_cosmo}. We see that, in contrast to the spherically symmetric case, here we have $\Lambda \geq 0$. Also, note that if the Hubble parameter, $H=\dot{a}/a$, is asymptotically constant, then $\bar{\Lambda}>0$, and $\alpha/a$ and $\dot{\alpha}/a$ go to zero as well as $\Lambda - \bar{\Lambda} \to 0$.

\begin{figure}[ht]
     \includegraphics[width=0.9\columnwidth]{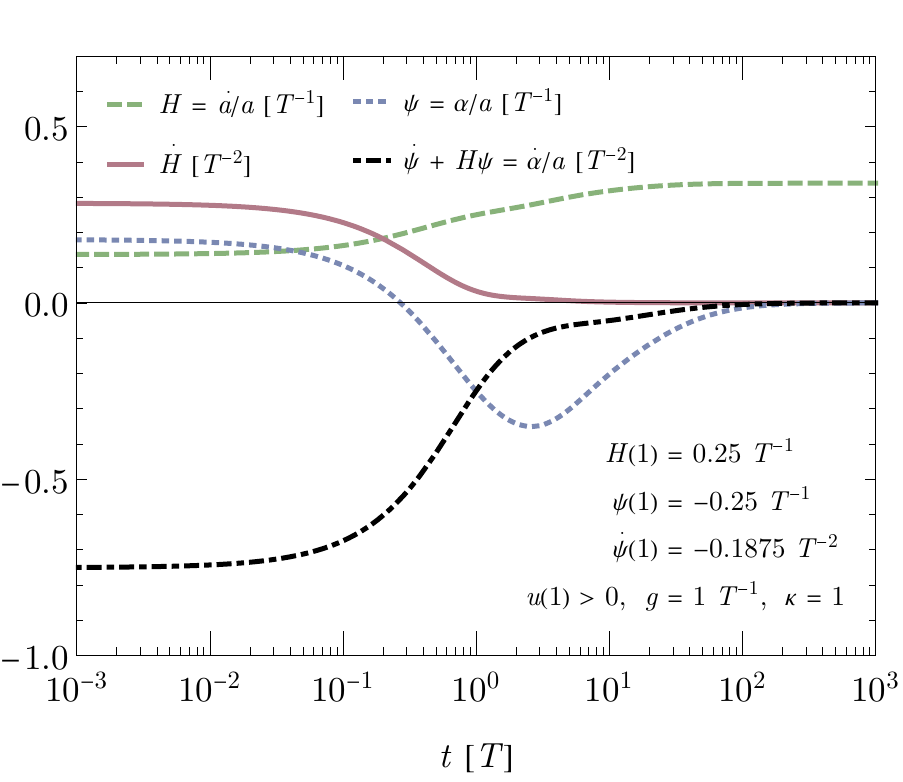}
    \caption{The evolution of scaled quantities $H$, $\dot{H}$, $\psi=\alpha/a$, and $\dot{\psi}+H\psi = \dot{\alpha}/a$ in the SU(2) cosmology model. All quantities are written in terms of $T$, the age of the Universe where the initial values are defined. With these initial conditions, $\bar{\Lambda} \approx 0.346\; T^{-2}$.
    }
    \label{fig:cosmological_parameters}
\end{figure}

The solution for a given set of initial conditions is shown in \figref{fig:cosmological_parameters}, which exhibits many of the properties already discussed. During the period where $\psi=\alpha/a$ is dynamical, $H$ undergoes an increase and eventually it levels off as $\psi$ goes to 0. This behavior may have interesting connections to late time dynamical dark energy models.

Note that due to the terms in parentheses in \eqref{eq:alphadotdot}, $\ddot{\alpha}$ is guaranteed to have a pole for some combination of parameters as long as $u$ is negative, which leads to singular behavior in the solutions. These poles lead to interesting behavior which cannot be solved by the numerical techniques used here.

\begin{figure}[ht]
     \includegraphics[width=0.9\columnwidth]{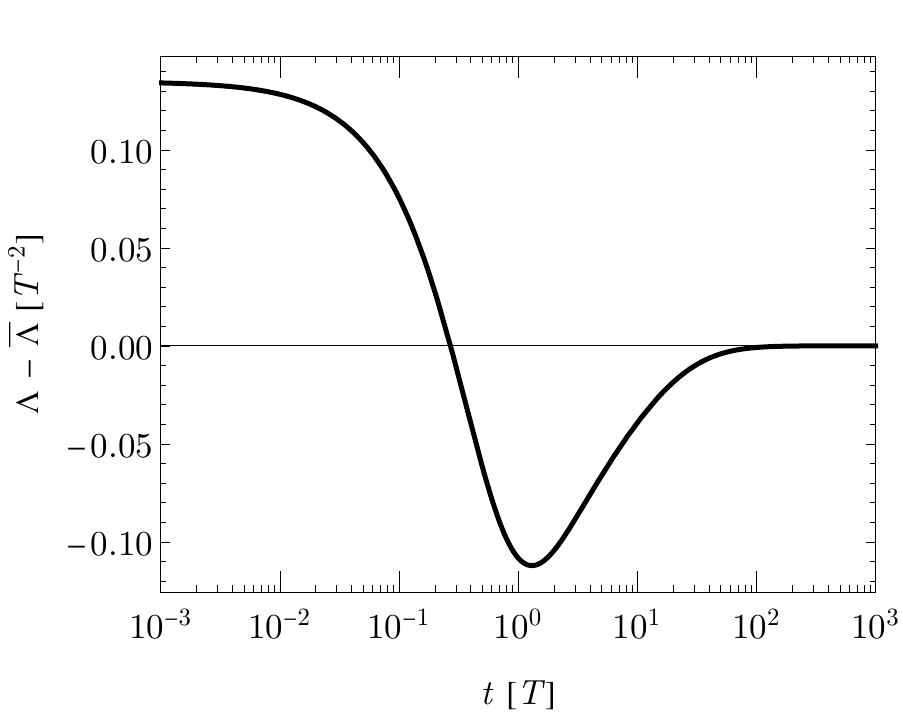}
    \caption{The evolution of $\Lambda-\bar{\Lambda}$ in the SU(2) cosmology model for the same initial conditions used in \figref{fig:cosmological_parameters}.
    }
    \label{fig:cosmological_lambda}
\end{figure}

We also show the evolution of $\Lambda-\bar{\Lambda}$ in \figref{fig:cosmological_lambda} using the same cosmology as \figref{fig:cosmological_parameters}. As expected from \figref{fig:cosmological_parameters}, the late time behavior is $\Lambda-\bar{\Lambda} \to 0$. While $H$ and $\Lambda$ have different behaviors when $\Lambda$ is varying rapidly, it is noteworthy that at late times, when $\Lambda$ approaches $\bar\Lambda$ and varies slowly, $H$ approaches the limiting value $(\bar\Lambda/3)^{1/2}$, as expected. An interesting behavior is that the early value of $\Lambda$ is greater than its late time value, even though $H$ increases with time. Note that the zeroes of $\Lambda - \bar\Lambda$ and $\psi$ coincide. In this particular example, $\Lambda$ varies by about a factor of two when comparing its minimum and maximum values.

Solutions also exist where $H$ is initially negative and changes sign, however these solutions run into pathologies at early times, and because we have relied on numerical methods, it is not clear whether these pathologies represent physical singularities or numerical artifacts. We hope to develop analytic methods for studying these cosmologies that can give us further insight into the early-time behavior of these bouncing cosmologies in future work.

\section{Discussion}
\label{sec:conclusion}
The theory we have presented in this article has some promising features: it is a minimal extension of Einstein gravity coupled to a topological field theory (BF theory), a system which has already been shown to resolve aspects of the cosmological constant problem; it predicts the existence of black holes that generalize the standard (A)dS-Schwarzschild solutions and approach their behavior asymptotically up to sub-leading corrections, and that (at least in the case when $\Lambda_0 - \bar\Lambda > 0$) each possess an equal unit ``charge," like fundamental particles; and it admits a simple homogeneous, isotropic cosmological model, which our numerical investigations have shown leads naturally to an effective cosmological constant that approaches a true constant at late times. 

While these preliminary investigations have uncovered some interesting features, more remains to be done in order to test the viability of the model. First, while the exact black hole solutions we have presented here are mathematically interesting in their own right, it remains to be seen what physically observable signatures might distinguish these black holes from those predicted by Einstein gravity, and whether those distinctions are supported or ruled out by astrophysical measurements. It would be interesting to subject this model to rotating black holes and see whether superradiance occurs and if it does, how it differs from general relativity or other modifications of gravity, such as dynamical Chern-Simons Gravity\cite{super,dcs} Finally, as we mentioned briefly in the text, the black hole solution we have focused on in this work is associated with a particular choice of a root of the cubic equation \eqref{cubic}. Using numerical methods, it can be seen that this solution is the only one that is well-defined for all positive values of the radial coordinate $r$. However, other solutions exist that are real for finite ranges $r \in (0, r_{max}]$. While it seems unlikely that such a solution could describe an astrophysical black hole, it would nevertheless be nice to know more about these solutions and understand their implications for a Birkhoff theorem of this model, or lack thereof, and to understand their behavior in the GR limit. 

For simplicity, we have not included any standard model matter in our homogeneous, isotropic model. It would be interesting to see which of the qualitative features we have observed in our numerical calculations persist in a more sophisticated analysis including matter and radiation, and in particular, whether the effective cosmological constant still asymptotes to a true constant at late times in that case.

 \section*{Acknowledgements}

The authors thank  Evan McDonough for insightful comments on an early draft of this work, and Leah Jenks, Konstantinos Koutrolikos, and Brad Marston for helpful discussions.

\appendix
\section{Spatial Integral Solutions} \label{ap:integration}
In this appendix, we provide a short derivation of the integrals shown in equations \eqref{eq:integral_1} and \eqref{eq:integral_2}, and provide some basic properties of equation \eqref{LpLb}. First, we begin by noting that equation \eqref{cubic} can be simplified dramatically by rewriting it in terms of dimensionless quantities
\begin{equation}
\frac{4}{27}q x^3 + x - 1 = 0
\label{eq:cubic_dimensionless}
\end{equation}
where
\begin{equation}
x=\frac{\Lambda-\bar{\Lambda}}{\Lambda_0-\bar{\Lambda}} \quad {\rm and} \quad q =\frac{27}{2}\frac{\left( \Lambda_0-\bar{\Lambda} \right)^3}{g^2 \kappa}r^4.
\label{eq:cubic_dimensionless_definitions}
\end{equation}
In these variables, \eqref{LpLb} takes on the simple form
\begin{equation}
x(q) = \frac{3}{2}\frac{\left( \sqrt{q^3+q^4}+q^2\right)^{2/3}-q}{q \left( \sqrt{q^3+q^4}+q^2\right)^{1/3}}.
\label{eq:cubic_sol_simple}
\end{equation}
Expansions of \eqref{eq:cubic_sol_simple} at extreme $r$ produce
\begin{align}
    \intertext{$\mathrm{for} \quad r \ll r_c \quad \mathrm{and} \quad \Lambda_0-\bar{\Lambda}>0$}
    \Lambda-\bar{\Lambda} \approx & \, (\Lambda_0-\bar{\Lambda}) + \mathcal{O}(r^4), \\
    \intertext{$\mathrm{for} \quad r \ll r_c \quad \mathrm{and} \quad \Lambda_0-\bar{\Lambda}<0$}
    \Lambda-\bar{\Lambda} \approx & \, \left|\frac{g^2 \kappa}{2} \frac{1}{\Lambda_0-\bar{\Lambda}}\right|^{1/2}r^{-2} + \mathcal{O}(r^{0}), \\
    \intertext{$\mathrm{for} \quad r \gg r_c$}
    \Lambda-\bar{\Lambda} \approx & \, \left(\frac{g^2 \kappa}{2} \right)^{1/3}r^{-4/3} + \mathcal{O}(r^{-8/3}),
\end{align}
where $r_c = |g^2 \kappa/2 (\Lambda_0 - \bar{\Lambda})^3|^{1/4}$, which corresponds to the location where the two extreme solutions intersect. $\Lambda-\bar{\Lambda}$ is finite for all values except $r=0$ for $\Lambda_0-\bar{\Lambda}<0$ where it diverges as $r^{-2}$.

While \eqref{eq:cubic_dimensionless} produces a simple form for $x(q)$, it will be more convenient for the following derivation to redefine it as
\begin{equation}
p r^4 x^3 + x - 1 = 0
\label{eq:cubic_dimensionless_2}
\end{equation}
with $p=2(\Lambda_0 - \bar{\Lambda})^3/g^2\kappa$. This change is to explicitly observe the $r$ dependencies of $x$ as well as limiting numerical constants during intermediate steps.
Differentiating \eqref{eq:cubic_dimensionless_2}  implicitly with respect to $r$ and solving for $\frac{dx}{dr}$ gives
\begin{align}
    \frac{dx}{dr}=-\frac{4 p r^3 x^3}{1 + 3 p r^4 x^2}.
    \label{eq:cubic_dimensionless_diff_2}
\end{align}

Next, we rearrange equation \eqref{eq:cubic_dimensionless_2} to get
\begin{equation}
    r=\left(\frac{1-x}{p x^3}\right)^{1/4}=p^{-1/4} \left(\frac{1-x}{x^3}\right)^{1/4}
    \label{eq:cubic_dimensionless_r_of_q_and_x_2}.
\end{equation}
Note that the separation of the fractions into two powers in \eqref{eq:cubic_dimensionless_r_of_q_and_x_2} is possible because $x \leq 1$; and if $x < 0$, then $p<0$. Now, consider the integral and its alteration using equations \eqref{eq:cubic_dimensionless_diff_2}, \eqref{eq:cubic_dimensionless_r_of_q_and_x_2}, and the first equation of \eqref{eq:cubic_dimensionless_definitions}: 
\begin{align}
    \int r^m (\Lambda - & \bar{\Lambda})^n dr = p^{-\frac{m+1}{4}} \frac{(\Lambda_0 - \bar{\Lambda})^n}{4}  \nonumber \\
    \times & \int \left(\frac{1-x}{x^3}\right)^{\frac{m-3}{4}} x^n \left( \frac{2}{x^3} - \frac{3}{x^4}\right) dx  
\end{align}
where $m$ and $n$ are positive integers. Integrals of this form produce the hypergeometric functions, ${}_2 F_1(a,b;c;z)$ which reduce down to the solutions shown in \eqref{eq:I_1_exact} and \eqref{eq:I_2_exact}.

\bibliographystyle{JHEP}
\bibliography{bibo}
	
\end{document}